\documentclass[aps,preprint,tightenlines]{revtex4}

\usepackage[english]{babel}
\usepackage{times}

\usepackage[]{graphicx}
\graphicspath{{figures/}}
\DeclareGraphicsExtensions{.eps}
\usepackage[figuresright]{rotating}

\usepackage{amsmath}

\newcommand{\Xmax}{X_{\text{max}}}
\newcommand{\RhoS}{\rho_{\text{s}}}
\newcommand{\RhoM}{\rho_{\mu}}
\newcommand{\Rms}{R_{\text{m.s.}}}

\begin{document}

\title{Recent results from Yakutsk experiment: development of EAS, energy
spectrum and primary particle mass composition in the energy region of
$10^{15}-10^{19}$\,eV}

\author{S.~P.~Knurenko}\email[]{s.p.knurenko@ikfia.ysn.ru}
\author{A.~A.~Ivanov}
\author{M.~I.~Pravdin}
\author{A.~V.~Sabourov}
\author{I.~Ye.~Sleptsov}
\affiliation{Yu.\,G.\,Shafer Institute for Cosmophysical Research and
Aeronomy,\\31 Lenin Ave., 677980 Yakutsk, Russia}

\begin{abstract}
  Experimental data obtained at the Yakutsk array after the modernization
  in 1993 are analyzed. The characteristics of EAS longitudinal and radial
  development found from the charged particle flux and EAS \v{C}erenkov
  light registered at the Yakutsk complex array are presented.  The energy
  spectrum of EAS obtained from \v{C}erenkov light and an estimate of the
  PCR mass composition are presented.
\end{abstract}

\maketitle

\section{INTRODUCTION}

The measurements of charged particles (electrons and muons with
$E_{\text{th}} \ge 1 \cdot \sec{\theta}$\,GeV) and \v{C}erenkov EAS
radiation are carried out at the Yakutsk EAS array during more than 35
years. After the modernization in 1993~\cite{b:1}, the Yakutsk array
significantly improves the measurement accuracy of main EAS characteristics
and increases a temp of statistics set in the energy range of $10^{17} -
5\cdot10^{18}$\,eV. It has come about through the increase of the number of
measurement stations and the decrease of separation between them.

A spectrum of energy, dissipated by primary cosmic rays (PCR) in extensive
air showers (EAS), have been obtained from these data and estimation of PCR
mass composition was made~\cite{b:2, b:3}. Another method for estimation of
primary particles mass composition will be presented hereinafter.

There is a perspective method for analysis of primary cosmic rays (PCR)
chemical composition based on conjoined analysis of longitudinal (cascade
curve) and lateral (structural functions of electron, muon and \v{C}erenkov
components) development of EAS. With such a complex approach to measurement
of shower characteristics here we have an opportunity of full
reconstruction of PCR mass composition using specific mathematical
techniques, for instance, inverse problem solving method~\cite{b:4, b:5,
b:6}, simplex method~\cite{b:7} and so on.

\section{LATERAL DISTRIBUTION OF DIFFERENT EAS COMPONENTS}

\subsection{Charged particles}

\begin{figure}[htb]
  \centering
  \includegraphics[width=0.60\textwidth, clip]{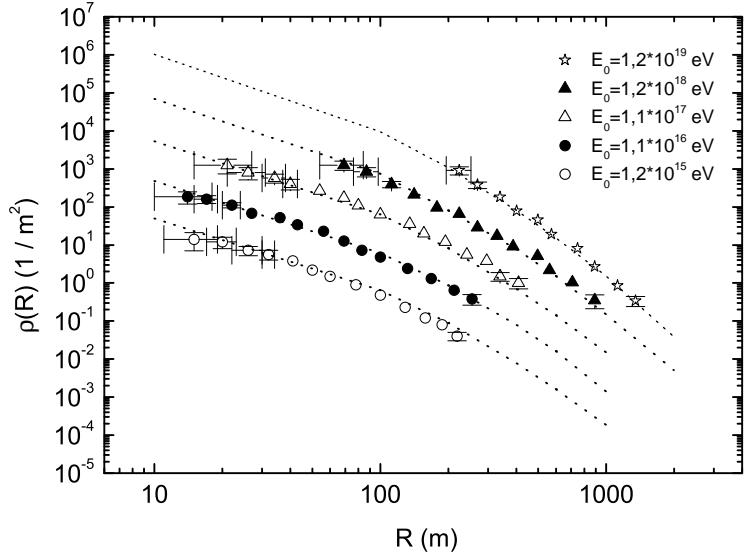}
  \caption{Average LDFs of charged particles measured at the Yakutsk EAS
  complex array.}
  \label{f:3a}
\end{figure}

Fig.~\ref{f:3a} shows the average lateral distribution functions (LDF) of
charged particles for $E_{0} \sim 10^{15} - 10^{19}$\,eV constructed by the
method used for the Yakutsk array~\cite{b:9}. From fig.~\ref{f:3a} it
follows that at $E_{0} \sim 10^{15} - 10^{17}$\,eV LDF’s are well measured
in the distance interval of $15-400$\,m from a shower core, whereas at
$E_{0} \ge 10^{17}$\,eV they are only measured in the distance interval of
$50-1300$\,m. The curves are the approximation $\rho(R)$ by the function
(\ref{eq:2}) from the work~\cite{b:10}:
\begin{equation}
    \rho(R) = \frac{N_{\text{s}}}{2 \pi R^{2}_{\text{m.s}}}
    \cdot \left(\frac{r}{\Rms}\right)^{-1.2} \times
    \left(1 + \frac{r}{\Rms}\right)^{-3.33}
    \cdot \left[1 + \left(\frac{r}{10 \Rms}
    \right)^{2}\right]^{-0,6}\text{,}
  \label{eq:2}
\end{equation}
where $N_{\text{s}}$ is the total number of charged particles at
observation level, $\Rms$ is a mean square radius of LDF of
charged particles.  It is seen from fig.~\ref{f:4a}, \ref{f:4b} that the
function (\ref{eq:2}) describes well experimental data both at $E_{0} \sim
10^{15}-10^{17}$\,eV and in the showers at $E_{0} \ge 10^{17}$\,eV.

\begin{figure}[htb]
  \centering
  \includegraphics[width=0.60\textwidth, clip]{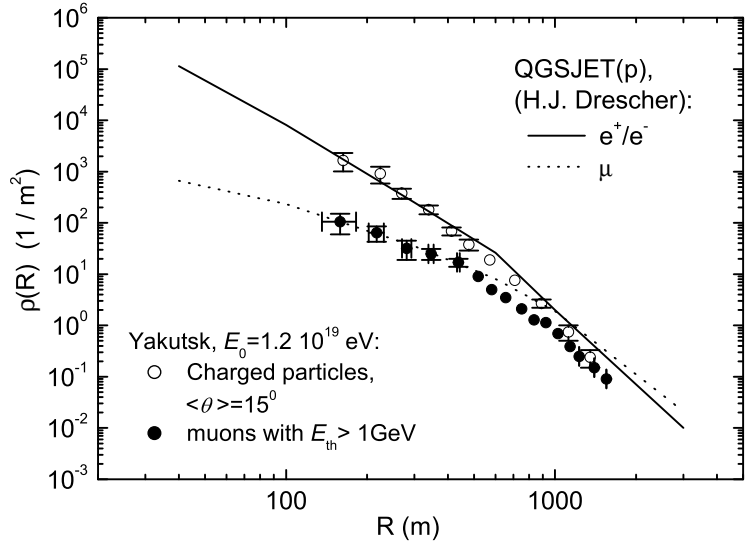}
  \caption{Lateral distributions of charged particles and muons. Circles
  are data from the Yakutsk experiment. Curves~--- from QGSJET
  model calculations for the showers generated by the primary proton.}
  \label{f:3b}
\end{figure}

Thus, the function (\ref{eq:2}) can be used to determine $N_{\text{s}}$ in
the showers at the highest energies where there are no measurements of
particle density at small distances.

On fig.~\ref{f:3b} a comparison is presented between the lateral
distributions of charged particles and muons with $E_{\text{th}} \ge
1$\,GeV and calculation result from work~\cite{b:11}. In the
work~\cite{b:11}, a hybrid scheme of EAS simulation is used together with
QGSJET01 model. As it follows from fig.~\ref{f:3b}, calculations give more
slope function as for charged particles, so for muons at distances more
than $600$\,m from the shower axis. If calculated value of $\rho(600)$
(QGSJET01 model) is used for shower energy estimation together with
model-independent method for energy estimation used in Yakutsk
experiment~\cite{b:12} then the energy estimated with QGSJET01 would be
underestimated by the value equal to the difference of densities
$\rho(600)$ (see fig.~\ref{f:3b}).

\subsection{\v{C}erenkov light}

\begin{figure}[htb]
  \centering
  \includegraphics[width=0.60\textwidth, clip]{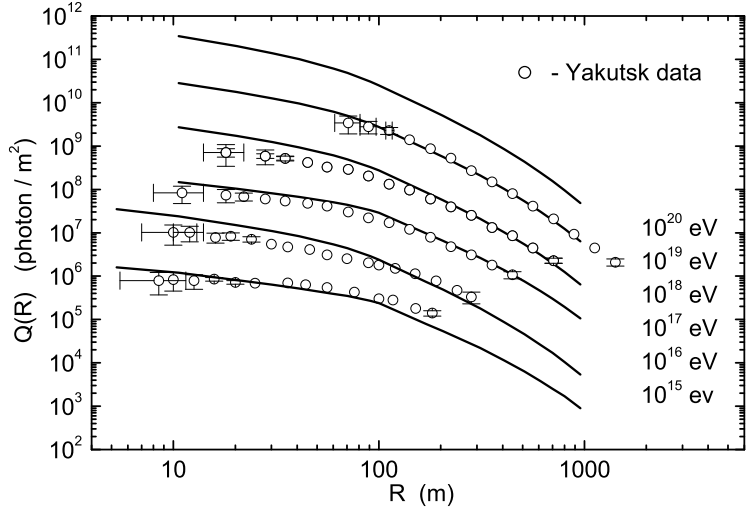}
  \caption{The lateral distribution of EAS \v{C}erenkov light. Curves~---
  QGSJET (Dedenko et al.).}
  \label{f:4a}
\end{figure}

\v{C}erenkov light measurements at the Yakutsk array last for more than 35
years. Since the year 1993 there are 50 operating \v{C}erenkov detectors
with receiving area of photocathode $176$ and $530$\,cm$^{2}$. Observation
results for last years are presented on fig.~\ref{f:4a}. On the same figure
calculations are shown from the work~\cite{b:13} (Dedenko et al.) for
primary proton and zenith angle $\theta = 0^{\circ}$. In the work a 5-level
scheme for air shower generation is used. One can see a good correspondence
in experimental data at medium and large distances from the axis. At
distances $< 80$\,m measured flux of \v{C}erenkov light is less than
one following from calculations. This discrepancy could be explained as with
distinct mass composition so with different zenith angle. Experimental LDFs
of \v{C}erenkov light are given for $\theta = 17-18^{\circ}$.

\begin{figure}[htb]
  \centering
  \includegraphics[width=0.60\textwidth, clip]{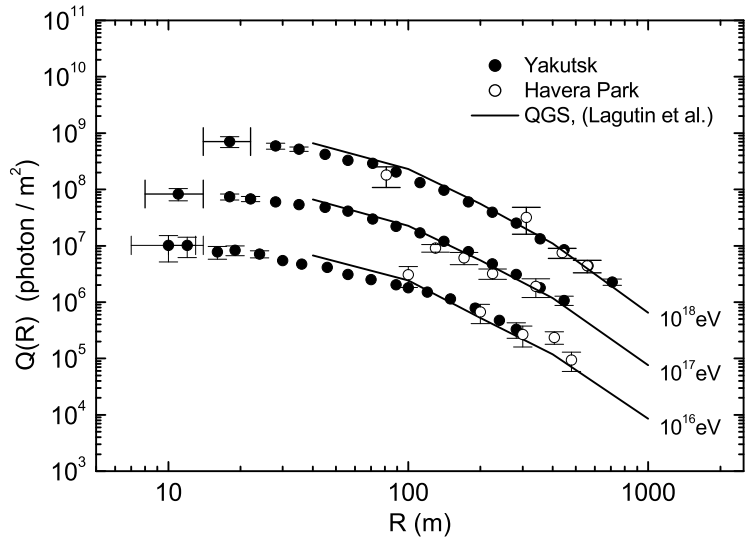}
  \caption{Comparison between the data from Yakutsk and Haverah Park.
  Curves~--- QGS (Lagutin et al.).}
  \label{f:4b}
\end{figure}

On fig.~\ref{f:4b} our results (circles) are given in comparison with the
data obtained by Haverah Park group~\cite{b:14}. The solid line is
calculation from work~\cite{b:15}. There is a good correspondence with the
experimental data.

\section{LONGITUDINAL DEVELOPMENT OF EAS}

\begin{figure}[htb]
  \centering
  \includegraphics[width=0.60\textwidth, clip]{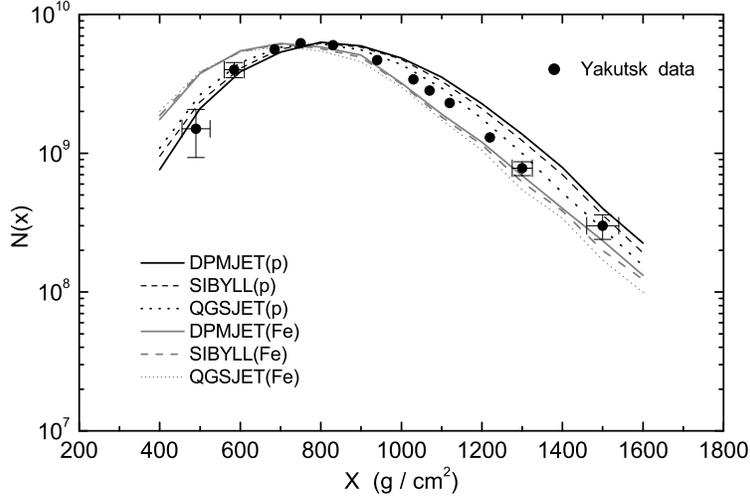}
  \caption{Average cascade curve of the shower development at $E_{0} =
  10^{19}$\,eV. Curves~--- different hadronic model calculations.}
  \label{f:5}
\end{figure}

Longitudinal development of high- and ultra-high energy air showers was
reconstructed from \v{C}erenkov light registered at the Yakutsk EAS array.
For this purpose a method proposed in work~\cite{b:6} was used. Results of
the reconstruction are shown on fig.~\ref{f:5} in comparison with
calculations obtained for different hadron interaction models. Calculations
have been performed for primary protons and iron nuclei. Experimental data
presented on fig.~\ref{f:5} are well described by models with fast
development like QGSJET in case of primary proton. It can be concluded that
primary mass composition is a mixture of nuclei with varying portion of
heavy nuclei. The most significant change is observed in the energy region
of $10^{16}-10^{17}$\,eV (experimental data tend towards heavy composition)
and above $3 \cdot 10^{18}$\,eV where mass composition is closer to proton.

\section{EMPIRICAL ESTIMATE OF EAS ENERGY AT THE YAKUTSK ARRAY}

The primary energy of a shower at the Yakutsk EAS array is calculated with
the expression

\begin{displaymath}
  E_{0} = E_{\text{ei}} + E_{\text{el}} + E_{\mu} + E_{\text{hi}} +
  E_{\mu\text{i}} + E_{\nu}\text{.}
\end{displaymath}
The energy scattered by electrons in the atmosphere above the observation
level is given by the expression
\begin{displaymath}
  E_{\text{ei}} = k(X, P_{\lambda}) \cdot F\text{,}
\end{displaymath}
Here, $F$ is total flux of \v{C}erenkov light from the EAS and $k(X,
P_{\lambda})$~--- is the coupling coefficient that represents the
transparency of the real atmosphere and character of the longitudinal
shower development, where $P_{\lambda}$ is spectral atmosphere transparency
(SAT), calculated during lidar measurements.

The energy of electrons at the observation level is calculated as
\begin{displaymath}
  E_{\text{el}} = 2.2 \cdot 10^{6} \cdot N_{\text{s}}(X_{0}) \cdot
  \lambda_{\text{eff}}\text{,}
\end{displaymath}
where $N_{\text{s}}(X_{0})$ is the total number of charged particles at sea
level and $\lambda_{\text{eff}}$ is the absorption mean free path of shower
particles obtained from the correlation of the parameters
$N_{\text{s}}(X_{0})$ and $Q(R=400)$ at different zenith angles. Other
components are: $E_{\mu} = \varepsilon_{\mu} \cdot N_{\mu}$;
$E_{\mu\text{i}} = (0.12 \pm 0.09) \cdot E_{\mu}$; $E_{\text{hi}} = (5.6
\pm 2.2) \cdot E_{\mu} $; $E_{\nu} = (0.64 \pm 0.18) \cdot E_{\mu}$.

\begin{figure}[htb]
  \centering
  \includegraphics[width=0.60\textwidth, clip]{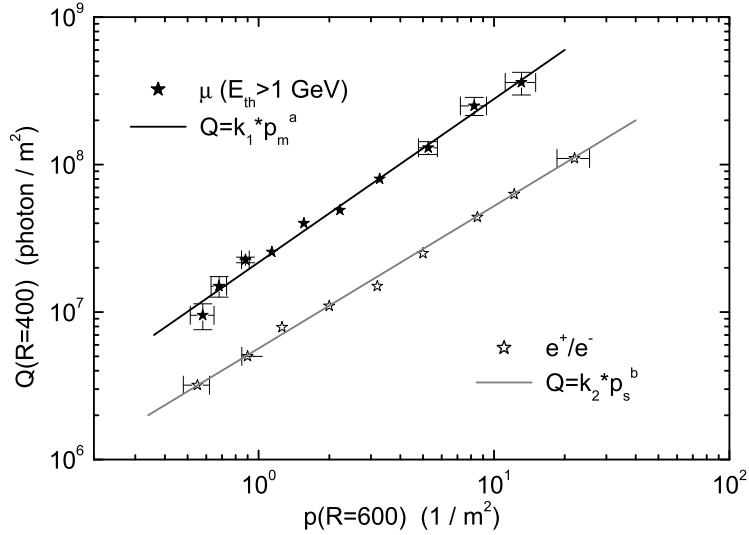}
  \caption{Estimation of the EAS energy by density $\RhoS(600)$,
  $\RhoM(600)$ and $Q(400)$.}
  \label{f:7}
\end{figure}

Using experimental data presented on fig.~\ref{f:7}, the following
expressions have been obtained at the Yakutsk array for energy estimation
by density of charged particles and muons with $E_{\text{th}} \ge 1$\,GeV:
\begin{eqnarray}
  \lg{E_{0}} & = & 17.68 + 0.98 \cdot \lg{\RhoS(600)}
  \label{eq:3} \\
  \lg{E_{0}} & = & 18.32 + 1.12 \cdot \lg{\RhoM(600)}
  \label{eq:4}
\end{eqnarray}

\section{ENERGY SPECTRUM IN THE ENERGY REGION OF $\sim
10^{15}-10^{20}$\,eV}

\begin{figure}[htb]
  \centering
  \includegraphics[width=0.60\textwidth, clip]{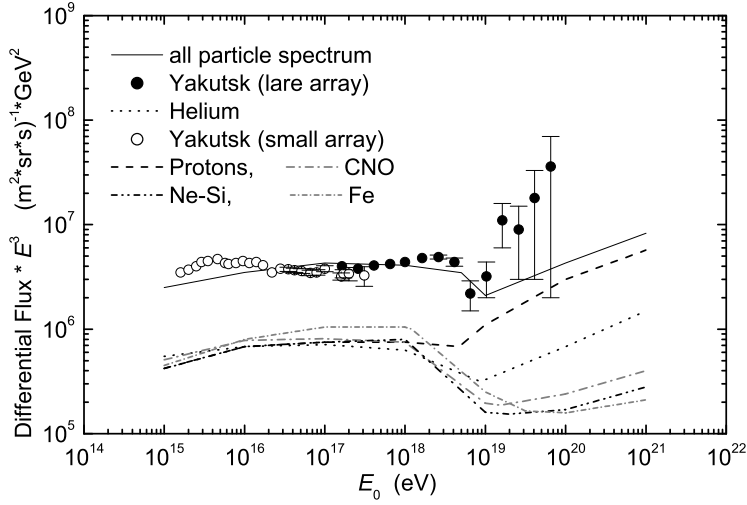}
  \caption{Spectra measured in Yakutsk using the air \v{C}erenkov light.
  Curves are results of anomalous diffusion model
  calculations~\cite{b:17}.}
  \label{f:8a}
\end{figure}

\begin{figure}[htb]
  \centering
  \includegraphics[width=0.60\textwidth, clip]{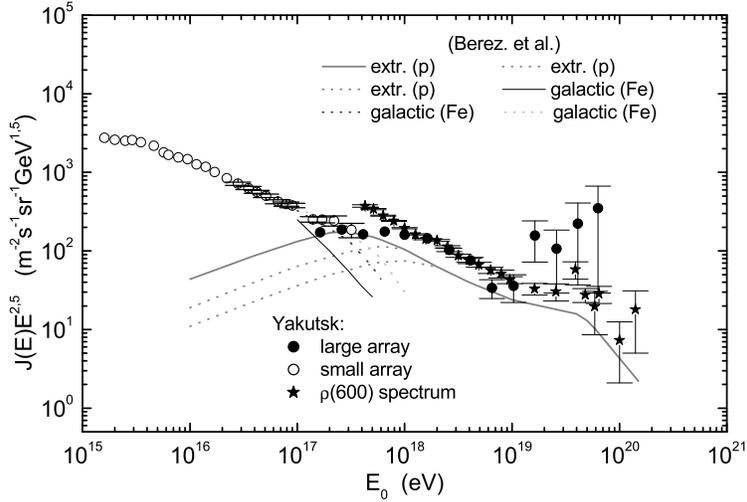}
  \caption{Calculated spectrum of extragalactic proton (solid curve) and of
  galactic iron spectra (dotted curves)~\cite{b:18} compared to all-particle
  spectrum from Yakutsk array.}
  \label{f:8b}
\end{figure}

In addition to charged particle surface detection, there is another
technique used at the Yakutsk array~--- the air \v{C}erenkov light
measurement, which can be used to draw out the cosmic ray spectrum in
independent way~\cite{b:16}. The spectrum covers wide energy region
$10^{15} - 10^{20}$\,eV. On fig.~\ref{f:8a}, fig.~\ref{f:8b} \v{C}erenkov
spectra are compared to calculations with different models of cosmic ray
propagation in the Universe~\cite{b:17, b:18}. It follows from
fig.~\ref{f:8a} and fig.~\ref{f:8b} that galactic model describes well our
spectrum in the region of $10^{15}-10^{18}$\,eV~\cite{b:17} and
metagalactic model~\cite{b:18}~--- above $10^{18}$\,eV. As it is seen from
fig.~\ref{f:8b}, in the region of $10^{17}-10^{18}$\,eV a boundary of
transition from galactic cosmic rays to metagalactic possibly exists. This
hypothesis requires further research.

\section{MASS COMPOSITION OF PRIMARY COSMIC RAYS}

To interpret Yakutsk experimental data a set of $\Xmax$ and $\RhoS(600)$
calculated values obtained with {\tt CORSIKA} simulation code (v.6.0,
QGSJET model) was used.  Calculations were carried for five primary nuclei
($\text{p}$, $\text{He}$, $\text{C}$, $\text{Si}$, $\text{Fe}$) at three
primary energy values $10^{17}$, $10^{18}$, $10^{19}$\,eV~\cite{b:19}. In
the work, two-dimensional probability densities $F(\Xmax, \rho(600))$ were
used with preliminary procedure of standardization of the experimental data
over the whole $(\Xmax, \RhoS(600))$ data set. In the numerical
implementation of this method, variables $\tau$ and $\rho$ were used
instead of $(\Xmax, \RhoS(600))$:
\begin{equation}
  \tau = \frac{\Xmax}{\sigma_{\text{x}}} - \left<
  \frac{\Xmax}{\sigma_{\text{x}}}
  \right>
\end{equation}
\begin{equation}
  \rho = \frac{\lg{\rho(600)}}{\sigma_{\lg{\rho(600)}}} - \left<
  \frac{\lg{\rho(600)}}{\sigma_{\lg{\rho(600)}}}
  \right>,
\end{equation}
where $\sigma_{\text{i}}$~--- is a mean square error.

For each of considered energy value and kind of primary nuclei (including
nuclei joint in groups $\text{p} + \text{He}$, $\text{C}$, $\text{Si} +
\text{Fe}$) probability distributions densities $f(\tau, \rho)$ were
constructed.  The intersection of $f(\tau, \rho)$ surfaces gives lines
$m_{1}$ and $m_{2}$ which optimally divide nuclei into ($\text{p} +
\text{He}$), $\text{C}$ and ($\text{Si} + \text{Fe}$) groups respectively.
The simulations showed that with dividing of data into three groups, the
effectiveness of nuclei group ($\text{p} + \text{He}$) to fall into the
zone~1 and of ($\text{Si} + \text{Fe}$) to fall into the zone~3 is up to
$90$\,\%. In the zone~2 a strong mixing between showers from different
primaries occurs and a portion of carbon is $\sim50$\,\% from all particles
in this zone.

The $\Xmax$ value characterizes a maximum of cascade development in
individual shower and $\rho(600)$ is the density of particles at the
observational level.

\begin{figure}[htb]
  \centering
  \includegraphics[width=0.60\textwidth, clip]{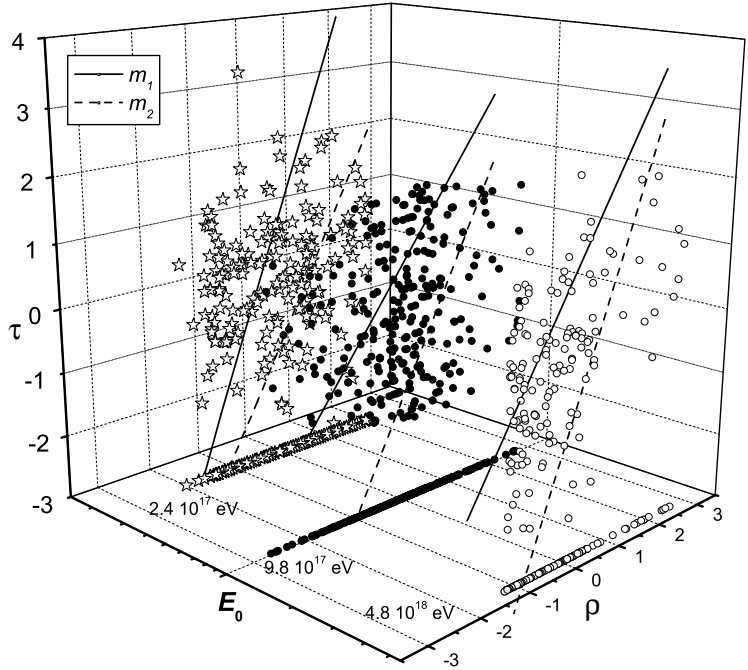}
  \caption{Standardized $\Xmax$ and $\rho_{600}$ experimental data
  at different energy values. $m_{1}$ is a borderline between nuclear
  groups ($\text{p} + \text{He}$) and $\text{C}$, $m_{2}$ is a borderline
  between nuclear groups $\text{C}$ and ($\text{Si} + \text{Fe}$).}
  \label{f:10}
\end{figure}

Fig.~\ref{f:10} presents the results of multicomponent analysis of $(\Xmax,
\rho(600))$ data set obtained at the Yakutsk EAS array. It is seen that
there is a correlation between the observed maximum of shower development
and the density of charged particles

The analysis was carried out for three values of energy, $2.4 \cdot
10^{17}$, $9.8 \cdot 10^{17}$ and $4.8 \cdot 10^{18}$\,eV. In this
notion, the point cloud represents standardized values, whose location
regions characterize zones directly connected with the mass number of a
primary particle.  Lines represent borders of such zones. In this case,
lines $m_{1}$ and $m_{2}$ optimally divide nuclei into groups ($\text{p} +
\text{He}$), $\text{C}$ and ($\text{Si} + \text{Fe}$). One can see from
fig.~\ref{f:10}, that the points are distributed over the zones
non-uniformly.

Distribution of statistics over the energy intervals draws attention. The
portion of ($\text{p} + \text{He}$) nuclei increases from $50$\,\% to
$53$\,\% and a portion of carbon nuclei~--- from $23$\,\% to $31$\,\%. At
the same time, the portion of heavy nuclei decreases from $27$\,\% to
$16$\,\% with growth of energy from $2.4 \cdot 10^{17}$\,eV to
$4.8 \cdot 10^{18}$\,eV. The error to recognize the nuclei for the energies
of $10^{17} - 10^{19}$\,eV does not exceed 30\,\%. Such a distribution of
nuclei in PCR does not contradict the conclusions about increase of the
portion of protons and helium nuclei in the limit energy region made in our
earlier works~\cite{b:20, b:21, b:22} where other methods were used.

\begin{figure}[htb]
  \centering
  \includegraphics[width=0.60\textwidth, clip]{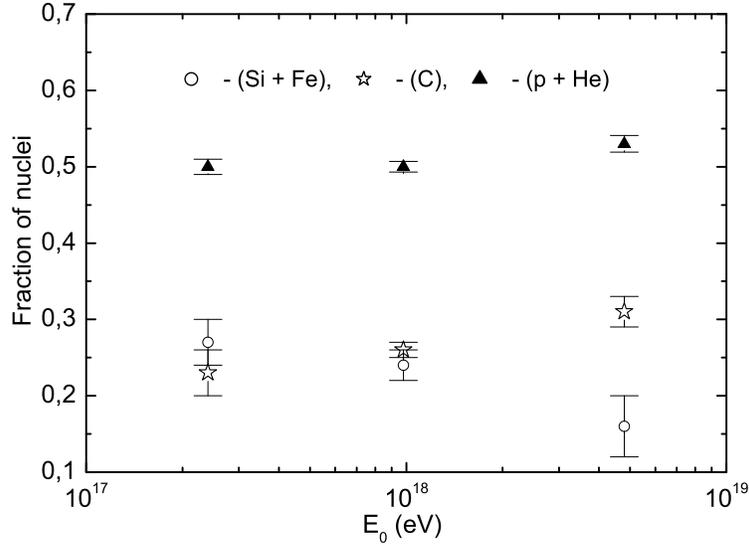}
  \caption{Fraction of the different nuclei in the energy range $E_{0} =
  10^{17} - 10^{19}$\,eV.}
  \label{f:11}
\end{figure}

\begin{figure}[htb]
  \centering
  \includegraphics[width=0.60\textwidth, clip]{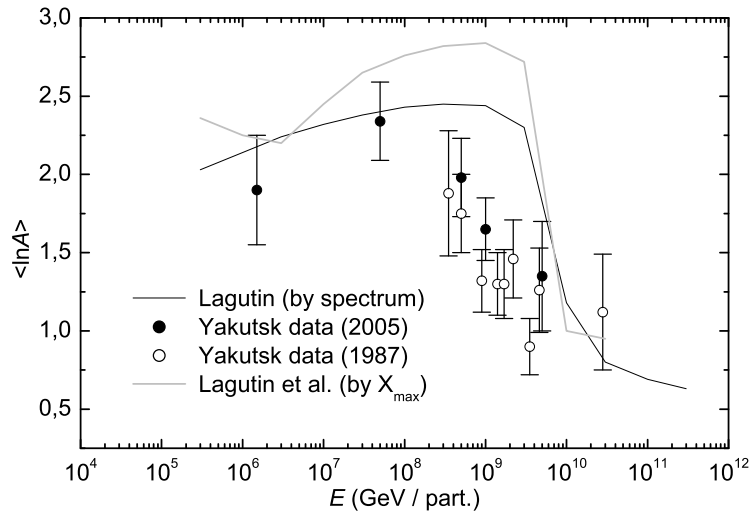}
  \caption{Mean mass numbers in cosmic rays of high- and ultra-high
  energies. Curves~--- calculations from anomalous diffusion model of
  cosmic rays propagation (Lagutin et al. 2004).}
  \label{f:12}
\end{figure}

On fig.~\ref{f:11} a portion of light, medium and heavy nuclei obtained
in our analysis is shown. It is seen from the figure that at $E_{0} \ge 4
\cdot 10^{18}$\,eV portion of heavy nuclei decreases. From fig.~\ref{f:12}
it follows that in the energy region $10^{16} - 10^{17}$\,eV $\left< \ln{A}
\right>$ value has its maximum value and after $5 \cdot 10^{17}$\,eV starts
decreasing. Such a change in mass composition in the energy region of
$10^{16} - 10^{17}$\,eV might be caused by modernized diffusion mechanism
of cosmic rays propagation in the Galaxy~\cite{b:17} and possible
influence upon the spectrum by lighter extragalactic component arriving
from beyond the Galaxy~\cite{b:18}.

\begin{acknowledgements}
  The work was partly financially supported by RFBR, grant
  No.\,06--02--16973--a, grant No.\,05--08--50045--a, grant
  NSh--7514.2002.2 and INTAS grant No.\,03--51--5112.
\end{acknowledgements}


\end{document}